\documentclass{article}

\usepackage{setspace}
\usepackage{spconf}
\usepackage{stmaryrd}
\usepackage{amssymb}
\usepackage{mathrsfs}
\usepackage{amsfonts}
\usepackage[dvips,final]{graphicx}
\usepackage[dvips]{color}
\usepackage{epsfig,amsmath}

\begin{document}
\title{Compressive Sensing for  MIMO Radar }\name{Yao Yu, Athina P.
Petropulu and H. Vincent Poor$^+$ \thanks{This work was supported in part by
the Office of Naval Research under Grant ONR-N-00014-07-1-0500}}
\address{Electrical \& Computer Engineering Department, Drexel University\\
$^+$ School of Engineering and Applied Science, Princeton
University}\maketitle

\begin{abstract}
{\small

Multiple-input multiple-output (MIMO) radar systems have been shown to achieve superior resolution as compared to traditional radar systems with the same number of transmit and receive antennas. This paper
considers a distributed MIMO radar scenario, in which each
transmit element is a node in a wireless network, and investigates
the use of compressive sampling  for direction-of-arrival (DOA)
estimation. According to the theory of compressive sampling, a
signal that is sparse in some domain can be recovered based on far
fewer samples than required by the Nyquist sampling theorem. The DOA
of targets  form a sparse vector in the angle space, and therefore,
compressive sampling can be applied for DOA estimation. The proposed
approach achieves the superior resolution of MIMO radar with far
fewer samples than other approaches. This is particularly useful in
a distributed scenario, in which the results at each receive node need
to be transmitted to a fusion center for further processing.
}\\


\end{abstract}
\section{Introduction}
Unlike a conventional transmit beamforming radar system that uses
highly correlated waveforms, a  multiple-input multiple-output (MIMO) radar system transmits  multiple independent waveforms
via its antennas \cite{Fishler:04}-\cite{Chen:08}. A MIMO radar system is
advantageous in two different scenarios \cite{Chen:08}. In the first
one, the transmit antennas are located far apart from each other relative
 to their distance to the target. The MIMO radar system
transmits independent probing signals from decorrelated transmitters
through different paths, and thus each waveform carries independent
information about the target. Therefore, the MIMO radar system can
reduce the target radar cross sections (RCS) scintillations and
provide spatial diversity. In the second scenario,
  a MIMO radar is equipped with $M_t$ transmit and $M_r$
receive antennas that are close to each other relative to the
target, so that the RCS does not vary between the different paths.
In this scenario,
 the phase differences induced by transmit
and receive antennas can  be exploited to form a long virtual array
with $M_tM_r$ elements. This enables the MIMO radar system to achieve
superior spatial resolution as compared  to a traditional radar
system. In this paper we consider  the second scenario.

Compressive sensing (CS) has  received considerable attention recently,
and has been applied successfully in diverse fields, e.g., image
processing \cite{Romberg:08} and wireless communications
\cite{Bajwa:06}. The theory of CS states that a $K$-sparse signal
$\mathbf{x}$ of length $N$ can be recovered exactly from
$\mathcal{O}(K\log N)$ measurements with high probability  via
linear programming. Let $\Psi$ denote the basis matrix that spans
this sparse space, and $\Phi$ a measurement matrix. The  convex
optimization problem arising from CS is formulated as follows:
\begin{eqnarray}
\min\|\mathbf{s}\|_1,\ \  \text{subject}\ \text{to}\ {\bf
y}=\Phi{\bf x} =\Phi\Psi{\bf s}
\end{eqnarray}
where $\mathbf{s}$ is a sparse vector with K principal elements and
the remaining elements can be ignored;  $\Phi$ is an $M\times N$
matrix incoherent with $\Psi$ and $M\ll N$.

The application of compressive sensing to a radar system was
investigated in \cite{Baraniuk:071}, \cite{Gurbuz:07} and
\cite{Herman:08}. In \cite{Baraniuk:071}, it was  demonstrated that
the CS method can eliminate the need for match filtering at the
receiver and has the potential to reduce the required sampling rate.
In the context of Ground Penetrating Radar (GPR), \cite{Gurbuz:07}
presented a CS data acquisition and imaging algorithm  that by
exploiting the sparsity of targets in the spatial space  can
generate sharper target space images with much less CS measurements
than the standard backprojection methods. Also the sparsity of
targets in the time-frequency plane was exploited for radar in
\cite{Herman:08}. In  the context of communication, \cite{Gurbuz:08}
proposed the direction of arrival estimation (DOA) estimation using
CS. In \cite{Gurbuz:08}, the basis matrix $\Psi$ is formed by the
discretization of the angle space. Since the signal sources were
assumed to be unknown, the basis matrix was approximated based on
the signal received by a reference vector. That signal would have to
be sampled at a very high rate  in order to construct a good basis
matrix.

In this paper, we extend the idea of
\cite{Baraniuk:071}-\cite{Gurbuz:08} to  the problem of  DOA
estimation  for MIMO radar. Since the number of targets is typically
smaller than the number of snapshots that can be obtained, DOA
estimation  can be formulated as the recovery of a sparse vector
using CS.  Unlike the scenario considered in \cite{Gurbuz:08}, in
MIMO radar the transmitted waveforms are known at each  receive
antennas, so that each receive antenna can construct the basis
matrix locally,  without the knowledge of the received signal at
other antennas. Further, radar systems often suffer from
interference due to jammers. Jammer suppression is investigated here
by exploiting the uncorrelatedness of  the transmitted waveforms
with the jammer signal in order to design the measurement matrix. We
provide analytical expressions for the signal-to-jammer ratio (SJR)
for the proposed approach. We also provide simulation results to
show that the proposed approach can accomplish super-resolution in
MIMO radar systems by using far fewer samples than existing methods,
such as Capon, amplitude and phase estimation (APES) and generalized
likelihood ratio test (GLRT) \cite{Xu:06}. This is very significant
in a distributed scenario, in which the receive nodes would need to
transmit the locally obtained information to a fusion center. For
such systems, we show that the proposed approach can enable each
node to obtain a good DOA estimate independently. Further, it
requires much less information to be transmitted to a fusion center,
thus enabling savings in terms of transmission energy.

\section{Signal Model for MIMO Radar}
We consider a MIMO radar system with $M_t$ transmit antennas and
$M_r$ receive antennas.  For
simplicity, we assume that targets and antennas   all lie in the same
plane. Let us denote the locations in rectangular
coordinates of the $i$-th transmit and receive
antenna by $(x^t_i, y^t_i)$ and $(x^r_i, y^r_i)$, respectively.
We assume that all transmit and receive
node locations relative to some reference point are known to each node
in the network. In a clustered system this information pattern may be achieved via a
beacon from the cluster-head \cite{Ochiai}.
The location of the $k$-th target is denoted by $(d_k,\theta_k)$,
where $d_k$ is the distance between this target and  the origin,
and $\theta_k$ is the azimuthal angle, which is the unknown
parameter to be estimated in this paper.

Under the far-field
assumption
 $d_{k} \gg \sqrt{(x^t_i)^2+(y^t_i)^2}$ and $d_{k} \gg\sqrt{(x^r_i)^2+
 (y^r_i)^2}$, the distance between the $i$th transmit/receive
 antenna and the $k$-th target
 $d^t_{ik}$/$d^r_{ik}$ can be approximated as
$d^{t/r}_{ik} \approx d_k - {\eta_i^{t/r}(\theta_k)} $, where
$\eta_i^{t/r}(\theta_k)={x^{t/r}_i \cos(\theta_k)+y^{t/r}_i
\sin(\theta_k)}$.

Let $x_i(n)$ denote the discrete-time waveform transmitted by the
$i$-th transmit antenna.  Assuming the transmitted waveforms are
narrowband and the propagation is non-dispersive, the received
baseband signal at the $k$-th target equals \cite{Li:07}
\begin{equation}
y_{k}(n)= \beta_k\sum_{i=1}^{M_t} x_i(n) e^{-j\frac{2\pi
}{\lambda}d^t_{ik} } =\beta_ke^{-j\frac{2\pi }{\lambda}d_k}{\bf
x}^T(n){\bf v}(\theta_k)
\end{equation}
for $k=1,\ldots, K$, where ${\bf x}(n)=[x_1(n),...,x_{M_t}(n)]^T$
and ${\bf
v}(\theta_k)=[e^{j\frac{2\pi}{\lambda}\eta^t_1(\theta_k)},...,e^{j\frac{2\pi}{\lambda}\eta^t_{M_t}(\theta_k)}]^T
$ is the steering vector.

Due to reflection by the target, the $l$-th antenna element
receives
\begin{equation}
z_{l}(n)=\sum_{k=1}^{K}e^{-j\frac{2\pi}{\lambda}d^r_{lk}}y_k(n)+\epsilon_{l}(n),\
l=1,\ldots, M_r
\end{equation}
where $\epsilon_{l}(n)$ represents independent and identically distributed  (i.i.d.) Gaussian noise with
variance $\sigma^2$.

On letting $L$ denote the number of snapshots, and placing $z_l(n), n=0,...,L-1$ in vector ${\bf z}_{l}$ we have
\begin{equation}
{\bf z}_{l}=
\sum_{k=1}^{K}e^{-j\frac{2\pi}{\lambda}(2d_k-
\eta^r_{l}(\theta_k))}\beta_k{{\bf X}}{\bf v}(\theta_k)+{\bf e}_l \label{4}
\end{equation}
where ${\bf y}_k=[y_k(0), \ldots, y_k(L-1)]^T$, ${\bf
e}_l=[\epsilon_{l}(0), \ldots, \epsilon_{l}(L-1)]^T$ and ${\bf X}=[
{\bf x}(0), \ldots, {\bf x}(L-1)]^T $.

By discretizing the angle space as
$\mathbf{a}=[\alpha_1,\ldots,\alpha_N]$, we can rewrite (\ref{4}) as
${\bf z}_{l}=\sum_{n=1}^{N}e^{j\frac{2\pi}{\lambda}
\eta^r_{l}(\alpha_n)}s_n{{\bf X}}{\bf v}(\alpha_n)+{\bf e}_l$,
%
where $N>>L$, and
\begin{equation*}
s_n = \left\{
\begin{array}{rl}
e^{-j\frac{4\pi}{\lambda}d_{k}}\beta_k &  \text{if there is target at}\ \alpha_n \\
0  & \text{otherwise}
\end{array} \right.  \ .
\end{equation*}

\section{Compressive Sensing for MIMO Radar}
Assuming that there exists a small number of targets, the DOAs are
sparse in the angle space, i.e., $\mathbf{s}=[s_1,\ldots,s_N]$ is a
sparse vector. A non-zero element with index $j$  in ${\bf s}$
indicates that there is a target at the angle $\alpha_j$. By CS
theory,  we can construct a basis matrix $\Psi_l$ for the $l$-th
antenna as
\begin{eqnarray}
\Psi_l=[e^{j\frac{2\pi}{\lambda} \eta^r_{l}(\alpha_1)}{{\bf X}}{\bf
v}(\alpha_1),\ldots,e^{j\frac{2\pi}{\lambda}
\eta^r_{l}(\alpha_N)}{{\bf X}}{\bf v}(\alpha_N)] \ .
\end{eqnarray}

  Ignoring the noise,  we have
$\mathbf{z}_l=\Psi_l\mathbf{s}$. Then we measure linear projections
of the received signal at the $l$-th antenna as
\begin{eqnarray} \label{receive_sig}{\bf
r}_l=\Phi_l\mathbf{z}_l={\Phi_l\Psi_l}\mathbf{s} \label{7}
\end{eqnarray}
where $\Phi_l$ is an $M\times L$ random Gaussian matrix which has
small correlation with ${\Psi}_l$. Placing  the output of  $N_r$
receive antennas, i.e., ${\bf r}_1,...,{\bf r}_{N_r}$, in vector ${\bf r}$ we have
\begin{equation}\label{cs}
{\bf r}={\Theta}\mathbf{s}, \quad 1\leq Nr\leq M_r \ .
\end{equation}
and the structure of $\Theta$ can be easily inferred based on (\ref{7}).
Therefore, we can recover $\mathbf{s}$ by applying the Dantzig selector
to the convex problem in (\ref{cs}) as in  \cite{Candes:07}:
\begin{eqnarray}\label{Dantzig}
\hat{{\bf s}}= \min\|{\bf s}\|_1\ \ \ s.t. \|{\Theta}^H({\bf
r}-\Theta{\bf s})\|_{\infty}<\mu.
\end{eqnarray}
According to \cite{Candes:07}, we can recover the sparse vector
${\bf s}$ with very high probability to select
$\mu=(1+t^{-1})\sqrt{2\log N\sigma^2}$, where $t$ is a positive
scalar.

\section{Performance Analysis in the presence of a jammer signal}

In this section, we analyze the effects of a jammer signal on the
performance of DOA estimation for MIMO radar using CS in terms of
the SJR. In the presence of a jammer signal, the received signal of
the $l$-th receive antenna is given by
\begin{eqnarray}
{\bf r}_{l}&=&\Phi_l\sum_{k=1}^{K}e^{-j\frac{2\pi}{\lambda}(2d_k-
\eta^r_{l}(\theta_k))}\beta_k{{\bf X}}{\bf
v}(\theta_k)\nonumber\\&&+\Phi_le^{-j\frac{2\pi}{\lambda}(d-
\eta^r_{l}(\theta))}\beta{{\bf b}}+\Phi_l{\bf e}_l \ .
\end{eqnarray}
The location of the jammer is denoted by  $(d,\theta)$, and $\beta$
and $\bf b$ denote respectively the reflection amplitude and
waveforms of this jammer. Since $\bf b$ is uncorrelated with the
transmitted waveforms $\bf X$, the effects of the  jammer signal are
similar to those of the addictive noise. Let ${\bf
A}_l=\Phi_l^H\Phi_l$. {Given the the transmit and receive node
locations, the average power of the desirable signal $P_s(l)$  over
the transmit waveforms is}
\begin{eqnarray}
P_s(l)&=&
E\{\sum_{k,k'=1}^{K}\underbrace{e^{j\frac{2\pi}{\lambda}[2(d_k-d_{k'})-(\eta^r_l(\theta_k)-\eta^r_l(\theta_{k'}))}}
_{\rho_{l}(k,{k'})}\beta_k^*\beta_{k'}\nonumber\\&&\times\underbrace{{\bf
v}^H(\theta_k) {{\bf X}}^H{\bf A}_l {\bf
X}{\bf v}(\theta_{k'})}_{R_{kk'}}\}=\underbrace{E\{\sum_{k=1}^{K}|\beta_k|^2 R_{kk}\}}_{C_1(l)}\nonumber\\
&&+ \underbrace{E\{\sum_{k\neq
k'}\rho_{l}(k,{k'})\beta_k^*\beta_{k'} R_{kk'}\}}_{C_2(l)}
\end{eqnarray}
where $C_1(l)$ is the dominant term and  $C_2(l)$ can be ignored
when the number of transmit antennas is sufficiently large.
Therefore, $P_s(l)$ can be approximated by $C_1(l)$. To simplify the
derivation of the SJR, we assume that the transmitted waveforms are
independently generated, orthogonal quadrature phase shift keyed
(QPSK) sequences with ${\bf X}^H{\bf X}={\bf I}_{M_t}$, ${\bf
b}^H{\bf b}=1$ and $\Phi_l\Phi_l^H={\bf I}_{M}$. Then $P_s(l)$ is
approximated by
\begin{eqnarray}
P_s(l)\approx\sum_{k=1}^K |\beta_k|^2\text{tr}({\bf
A}_l)M_t/L=\frac{MM_t}{L}\sum_{k=1}^K |\beta_k|^2 \ .
\end{eqnarray}

Similarly, {the average power of the jammer interference over the
jammer waveforms} is given by
\begin{eqnarray}
P_j(l)&=& E\{(e^{-j\frac{2\pi}{\lambda}(d-\eta^r_l(\theta))}
\beta)(e^{-j\frac{2\pi}{\lambda}(d-\eta^r_l(\theta))} \beta)^*\nonumber\\
&&\times{\bf b}^H{\bf A}_l{\bf b}\}=|\beta|^2M/L \ .
\end{eqnarray}

From these two expressions, the SJR becomes
\begin{eqnarray}
\text{SJR}=\frac{P_s(l)}{P_j(l)}\approx\frac{M_t\sum_{k=1}^K
|\beta_k|^2}{|\beta|^2} \ .
\end{eqnarray}

Since the jammer signal is uncorrelated with the transmitted signal,
the SJR can be improved by correlating the jammer signal with the
transmitted signal. Combining this with CS, the measurement matrix
in (\ref{receive_sig}) is modified as $\tilde{\Phi}_l=\Phi_l{\bf
X}^H$. Moreover, since  $\Phi_l$ is a Gaussian random matrix,
$\tilde{\Phi}_l$ is still Gaussian; therefore it satisfies the
restricted isometry property (RIP)  and is incoherent with
 $\Psi_l$, thus guaranteeing a stable solution to
(\ref{Dantzig}). Based on $\tilde{\Phi}_l$, the SJR can be obtained
as
\begin{eqnarray}\label{SJR}
\text{SJR}\approx\frac{L\sum_{k=1}^K |\beta_k|^2}{|\beta|^2} \ .
\end{eqnarray}
Generally, the SJR can be improved by a factor of $L/M_t$ using
$\tilde{\Phi}_l$ since $L\gg M_t$. (\ref{SJR}) indicates that the
increase in $L$ will improve the DOA estimates.  However, more
calculations are required by $\ell_1$-norm minimization due to the
increase in the size of the basis matrix, and the time duration of
the radar pulse needs to be longer as well.

The proposed method is especially advantageous in a distributed MIMO
radar system in which the  receive elements are randomly
distributed.  In particular, many fewer measurements are required to be
sent to the base station  or fusion center (FC) in this situation than are
need by  conventional methods. As simulation results show (see Section
\ref{simulation}), the proposed method  can yield  good performance
even using a single receive antenna. With a good initial estimate of
DOA, the receive nodes can  adaptively refine their estimates by
constructing  a higher resolution basis matrix $\Psi_l$ around that
DOA.
 Restricting the candidate angle space,
 may reduce the samples in the angle space that are required
for constructing the basis matrix, thus reducing the complexity of the
$\ell_1$ minimization step. On the other hand, the resolution of
target detection can be improved by taking the denser samples of the
angle space around the intimal DOA estimate.

\section{Simulation Results }\label{simulation}

In this section, we consider a MIMO radar system with the
transmit/receive antennas
 randomly distributed within a small area on a two-dimensional
(2-D) disk.  $M_t=30$ antennas transmit independent QPSK waveforms.
The carrier frequency is 8.62 GHz. A maximum of $L = 512$ snapshots
are considered at the receive node. The received signal is
 corrupted by  zero mean Gaussian noise.
 The SNR is
set to 20 dB. There are two targets located at
$\theta_1=-3\textordmasculine$ and $\theta_1=-2\textordmasculine$,
with reflection coefficients $\beta_k=1, k=1,2$. A jammer is located
at $0\textordmasculine$ with an unknown Gaussian random waveform and
with amplitude 10, i.e., 20 dB above the target reflection
coefficients
 $\beta_k$. We sample the angle space by increments of $0.2\textordmasculine$
from $-5\textordmasculine$ to $5\textordmasculine$, i.e., ${\bf
a}=[-5\textordmasculine,-4.8\textordmasculine,\ldots,4.8\textordmasculine,5\textordmasculine]$.
 We compare the performance of the proposed
method and three approaches \cite{Xu:06}, i.e., the Capon, APES and
GLRT techniques.

\begin{figure}[htbp]
  \centering
    \includegraphics[height=2.2in,width=3.6in,clip=true]{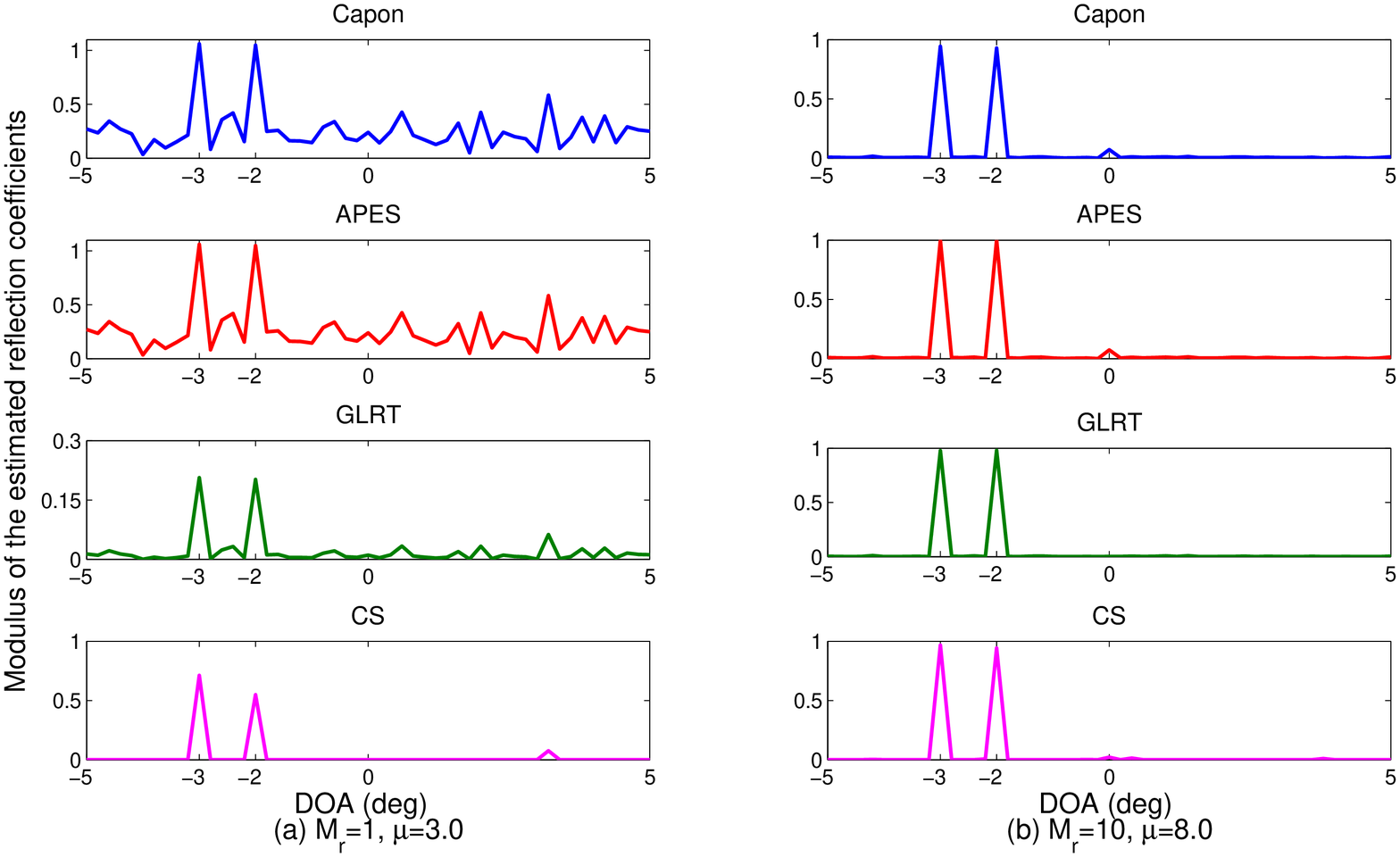}
 \caption{\scriptsize DOA estimation using (a) 1 receive antenna and (b) $10$ receive
 antennas. In both (a) and (b), the top three curves were obtained using $512$ snapshots.
 The bottom curve was obtained using $15$ snapshots only.
  }\label{veryclose}
\end{figure}

Figs. \ref{veryclose} shows the moduli of the estimated reflection
coefficients $\beta_k$, as functions of the azimuthal angle for
(a) $M_r=1$ and (b) $10$ receive antennas, respectively. In both (a) and (b),
the top three curves correspond to the azimuthal estimates obtained
via Capon, APES and GLRT, using $512$ snapshots. The bottom curve is
the result of the proposed approach, obtained using $15$ snapshots
only. One can  see that in the case of using only one receive node,
the presence of the two targets is clearly evident via the proposed
method based on $15$ snapshots only. The other methods produce
spurious peaks away from the target locations. When the measurements
of multiple receive nodes are used at a fusion center, the proposed
approach can yield similar performance to the other three methods.
However, the comparison methods would have to transmit to the fusion
center $512$ received samples each, while in the proposed approach,
each node would need to transmit $15$ samples each.

The threshold $\mu$ in
(\ref{Dantzig})  affects DOA estimation for the proposed method.
The increase in $\mu$ within a range can reduce the ripples of DOA
estimates at the non-target azimuth angles at the expense of
 the accuracy of the target-reflection-coefficient
estimates.  The increase in $\mu$ can also reduce the complexity of
(\ref{Dantzig}) because the constraint is  looser than that of
smaller $\mu$. If $\mu$ is too large, however, the $\ell_1$-norm
minimization does not work. In
Fig.\ref{veryclose}, a relatively large threshold, i.e.,  $\mu=3$, was used for the
single receive node case. As a result, the CS method yielded less accurate  estimates of the reflection coefficients magnitude
 than the Capon and GLRT, but with
 very few ripples.

\begin{figure}[htbp]
  \centering
    \includegraphics[height=2.0in,width=3.6in,clip=true]{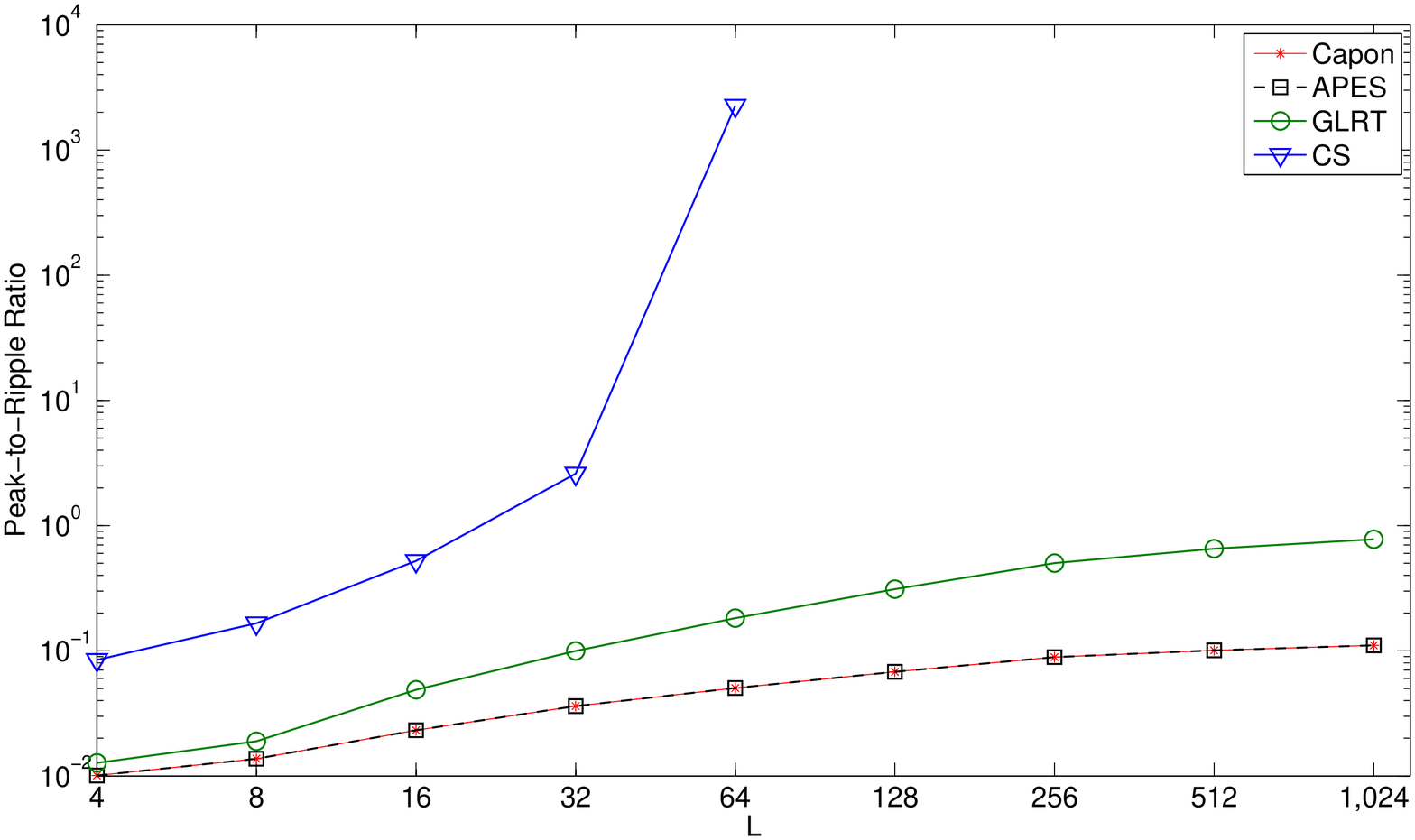}
 \caption{\footnotesize PRR using one receive antenna.
  }\label{compare_peak_ripple}
\end{figure}

 Fig. \ref{compare_peak_ripple} shows the effect of the number
of snapshots $L$ on the DOA estimates of the proposed method and the
other methods. Here, we  consider the case of one receive
antenna only. In order to quantify the performances of DOA
estimation, we define the ratio of the square amplitude  of the DOA
estimate at the  target azimuth angle to the sum of the square amplitude
of DOA estimates at other angles as the peak-to-ripple ratio (PRR).
As shown in Fig.  \ref{compare_peak_ripple} , although the increase
in $L$ can improve the PPR of these four methods, the increase is much
faster for  the CS method.

\section{Conclusion}
A compressive sensing method has been proposed to estimate the DOA
of targets for MIMO radar systems. The DOA of targets  can construct
a sparse vector in the angle space. Therefore, we can solve for this
sparse vector by $\ell_l$-norm minimization with many fewer samples
than  conventional methods, i.e. the Capon, APES and GLRT
techniques.  The proposed method  is superior to these conventional
methods when one receive antenna is active. If multiple receive
antennas are used, the proposed approach can yield similar
performance  to the other three methods, but by using far fewer
samples.\\

\centerline{\bf Acknowledgment} The authors would like to thank Dr.
Rabinder Madan of the Office of Naval Research for bringing the possibility of using compressive
sensing for angle-of-arrival estimation to their attention.

\bibliographystyle{IEEE}

\end{document}